\documentclass{article}

\usepackage{arxiv}
\usepackage{amsmath}
\usepackage[utf8]{inputenc} 
\usepackage[T1]{fontenc}    
\usepackage{hyperref}       
\usepackage{url}            
\usepackage{booktabs}       
\usepackage{amsfonts}       
\usepackage{nicefrac}       
\usepackage{microtype}      
\usepackage{cleveref}       
\usepackage{lipsum}         
\usepackage{graphicx}

\usepackage{natbib}
\usepackage{doi}

\title{A multi-year CONUS-wide analysis of lightning strikes to wind turbines}
\date{October 10, 2024}	

\author{\href{https://orcid.org/0000-0002-8095-4204}Ryan K.~Said \\ 
	Vaisala Inc.\\
	Louisville, CO 80027 \\
	\texttt{ryan.said@vaisala.com} \\
	\And
	Eric Grimit \\
	Vaisala Inc.\\
	Seattle, WA\\
  \And
	Martin Murphy \\
	Vaisala Inc.\\
	Louisville, CO\\
}

\renewcommand{\shorttitle}{Lightning strikes to turbines}

\hypersetup{
pdftitle={Ltg strikes to turbines},
pdfsubject={stats.AP},
pdfauthor={Ryan Said, Eric Grimit, Martin Murphy},
pdfkeywords={Lightning, lightning measurements, upward-initiated lightning, wind turbines}
}

\begin{document}
\maketitle


\begin{abstract}
  Lightning strikes to wind turbines (WTs) pose significant hazards and operational costs to the renewable wind industry. These strikes fall into two categories: downward cloud-to-ground (CG) strokes and upward discharges, which can be self-initiated or triggered by a nearby flash. The incidence of each type of strike depends on several factors, including the electrical structure of the thunderstorm and turbine height. The strike rates of CG strokes and triggered upward lightning can be normalized by the amount of local CG activity, where the constant of proportionality carries units of area and is often termed the collection area. This paper introduces a statistical analysis technique that uses lightning locating system (LLS) data to estimate the collection areas for downward and triggered upward lightning strikes to WTs. The technique includes a normalization method that addresses the confounding factor of neighboring WTs. This analysis method is applied to seven years of data from the National Lightning Detection Network$^\textrm{TM}$ and the US Wind Turbine Database to investigate the dependence of collection areas on blade tip height and peak current. The results are compared against estimates of collection areas derived from counts of LLS-detected CG strokes close to WTs.
\end{abstract}


\section{Introduction}

Renewable energy generation capacity from wind turbines (WTs) has steadily increased in the United States and internationally over the past 25 years. In 2022, wind energy accounted for $\sim$10\% of the total energy capacity in the US \citep{wiser2023land}. The growth in this capacity has been accompanied by advancements in WT technology, including developing turbines with larger hub heights and rotor diameters.  The average tip height, defined as the distance from the ground to the blade tip when it is vertical, has increased from less than 100 meters in 2000 to 164 meters in 2022. This increase in overall height and swept area has resulted in a threefold increase in capacity per turbine. The increased size also brings an additional hazard: the higher tip heights make them more susceptible to lightning strikes \citep{eriksson1987incidence}. Additional lightning strikes present both safety and operational risks. The powerful electrical currents and intense heat generated by lightning discharges can cause significant damage to turbine blades \citep{candela2014lightning} and electrical components \citep{rodrigues2010electromagnetic}. Lightning protection system (LPS) standards are based on observed distributions of the various electrical current parameters that cause damage. While the highest standards target protection against 99\% of lightning discharges across each parameter \cite{international2006protection}, more frequent lightning discharges imply greater probability of a particularly energetic discharge that exceeds the LPS tolerances \citet{international2019wind}. Furthering the risk, the electrical current may not attach to a designated contact point in the LPS, and LPS components can deteriorate over time after repeated exposure to electrical currents \citep{madsen2023wind}. 

Tall structures, including WTs, are prone to lightning strikes during thunderstorms or when low-altitude electrification occurs in the atmosphere. There are three categories of lightning strikes to tall structures, each influenced by distinct factors that control the incidence rate, and each having different electric current processes operating with their own time scales and characteristic amplitudes. The LPSs on WTs are designed according to the distributions of the magnitudes and time scales of each of these electrical current processes \citep{madsen2023wind}. The following paragraphs describe these categories and their salient characteristics. 

The first category is cloud-to-ground (CG) strokes, which are preferentially attracted to elevated points or structures. CG return strokes are preceded by a charged stepped leader. As the leader approaches the ground, it induces oppositely charged upward leaders due to the enhanced electric field \citep{rizk1994modeling}. When the downward leader connects with an upward leader, an upward-propagating return stroke process neutralizes the charge deposited along the channel, allowing electrical current to flow through the ground contact point. Elevated structures enhance the induced electric field, increasing the likelihood of developing an upward leader \citep{becerra2006simplified,saba2022close}. At the attachment point, the electrical current of downward CG lightning begins with a rapid rise to the peak over a few microseconds after the downward stepped leader makes contact with an upward connecting leader \citep{dwyer2014physics}. The peak current can range from a few to several hundred kiloamps \citep{berger1975parameters}. Following the peak, the current decays over tens of microseconds. Multiple return strokes may recur along the same channel, separated by tens to hundreds of milliseconds. Some downward CG strokes may also be followed by a continuing current (CC), which has a much lower amplitude\textemdash typically a few hundred amps\textemdash but persists for much longer than the current surge in a return stroke, ranging from a few to many hundreds of milliseconds. Due to its long duration, CC can deposit significant charge, leading to damage caused by heating. 

The second and third categories of lightning strikes involve upward lightning (UL) discharges, which impact tall structures above approximately 100 meters. All UL begins with an upward leader that develops in response to a strong electric field at the structure, though this electric field enhancement is not from an approaching stepped leader. \citet{wang2008observed} divided upward flashes into two categories based on whether another lightning discharge precedes the UL: triggered (T) and self-initiated (SI) upward lightning. We consider each of these types as their own category of lightning strikes to WTs since different factors control their occurrence rates. As the names suggest, triggered UL (TUL) follows a nearby triggering flash, whereas self-initiated UL (SIUL) occurs without prior lightning activity. In both cases, the electric fields at the top of the structure are sufficiently high to promote an upward-propagating leader. In TUL, leaders in the cloud or nearby lightning flashes induce a transient electric field change at the structure's tip. A large collection of studies has shown that the most common trigger is from negative leaders propagating overhead that originated from a nearby positive CG flash \citep{warner2012upward, saba2016upward, yuan2017characteristics, schumann2019triggering, zhu2019evolution}. However, a variety of triggering events have been observed, including leader propagation preceding a CG flash, in-cloud flashes, and even upward lightning from nearby towers \citep{candela2014lightning,schumann2019triggering,sunjerga2021initiation}. In contrast, SIUL develops without any triggering lightning flash. In this case the electric field enhancement comes from the proximity between the object tip to a charge layer in the atmosphere \citep{rizk1994modeling}. Therefore, SIUL typically occurs in cold months when the charge layer is at a low altitude \citep{warner2014synoptic, shindo2018lightning}.

Compared to CG lightning, UL tends to have lower peak currents, but a higher distribution of charge transfer \citep{berger1978blitzstrom}. Each UL discharge begins with an initial stage (IS) as an upward leader propagates towards the cloud \citep{miki2005initial,saba2016upward}. The IS lasts from a few tens to hundreds of milliseconds, and is characterized by an initial continuous current (ICC). Similar to the CC stage in downward CG lightning, the ICC has a low amplitude of a few hundred amps but can transfer significant total charge. The IS may also include current pulses, typically on the order of a few thousand amps \citep{miki2005initial}. The UL discharge may cease after the IS, or one or more return strokes may occur, propagating down the ionized channel left by the upward leader. These return strokes are phenomenologically similar to subsequent return strokes in downward lightning flashes, although they generally have lower peak current amplitudes.

This paper is concerned with characterizing the strike rates to WTs among these three categories (CG, TUL, and SIUL) of lightning. Historically, much of our understanding of strike rates to tall structures comes from instrumented towers \citep{eriksson1987incidence}. National scale lightning locating systems \citep{cummins2009overview} (LLSs) offer an alternative means to detect lightning strikes over a large geographic area. These systems detect the strong radio pulses emitted by CG return strokes and some cloud discharges. While the detection efficiency (DE) for CG flashes is typically over 95\% (e.g., \citep{mallick2014performance}), the DE for upward discharges is significantly lower. While the current surge associated with a return stroke generates a radio burst that can be reliably detected and located by a large-area LLS \citep{nag2015lightning}, the long time scales associated with CC/ICC render their radio emissions less detectable. Consequently, the fraction of detected upward discharges depends on the presence of return strokes following the IS or sufficiently strong pulses superimposed on the ICC.

Tower measurements in Europe have quantified the fraction of upward discharges detected by a precision LLS. Using in-situ current measurements from the Peissenberg tower in Germany (150 m height), \citet{paul2019performance} analyzed the fraction of different types of upward discharge current impulses detected by the EUCLID LLS \citep{schulz2016european}. Over the observational period, EUCLID detected 51 out of 199 (25.6\%) current pulses of all types. The breakdown of the detection efficiency by type reflects the magnitude of the radio impulse within the frequency range detectable by the LLS. Out of the 129 ICC and other non-return-stroke pulses, only 11 (8.5\%) were detected by EUCLID, whereas 40 out of the 49 return strokes (81.6\%) were detected. \citet{diendorfer2015lls} reported the EUCLID DE of UL at the flash level using measurements from the Gaisberg Tower in Austria. The DE for flashes with only an ICC and no appreciable superimposed impulses was zero, whereas the DE for upward flashes with superimposed impulses increased to 58\%. The DE for upward flashes that included return strokes was consistent with the DE for downward CG flashes, around 96\%. Both studies show that the overall upward flash detection depends on the distribution of flashes of each type. In the \citet{diendorfer2015lls} campaign, the distribution was 47\% ICC-only, 21\% with superimposed pulses, and 32\% with return strokes, resulting in a total upward flash-weighted DE of 43\%. \citet{watanabe2019characteristics} summarized several studies characterizing the fraction of UL with one or more return strokes, finding that the fraction ranges from 11\% to 50\%, with the majority of studies reporting around a quarter to a third of UL containing one or more return strokes.

While the DE of upward flashes is lower for large-scale LLSs compared to instrumented tower measurements, several researchers have nonetheless leveraged the extensive large-volume statistical data provided by LLSs to study the occurrence rates of lightning strikes to tall structures and WTs. \citet{smorgonskiy2013proportion} proposed a method to estimate the fraction of upward discharges based on the ground flash density near the structure and an assumption of the attraction radius $R_\textrm{a}$, described in Section \ref{sec:sr_summary}. Within this attraction radius, the total number of flashes includes upward flashes from the structure and downward flashes that strike either the structure or the nearby ground. If the attraction radius is known, the number of strikes to the structure can be estimated from the local ground flash density.  \citet{soula2019quantifying} examined 21 years of LLS data over a small area containing eight wind farms in southwestern France. The study found that the fraction of CG strokes within 200 m of each turbine increased by a factor of 3.11 after the installation of the WTs. This increase includes contributions from attracted CG strokes and the introduction of upward discharges from the WTs. 

The present paper introduces a statistical analysis technique that leverages CG return stroke data from a national-scale LLS to characterize WT strike rates from upward and downward discharges. This study sources lightning data from the National Lightning Detection Network$^\textrm{TM}$ (NLDN$^\textrm{TM}$) \citep{cummins2009overview}, which, like EUCLID, detects the vast majority of CG flashes within its region of coverage \citep{mallick2014performance}. Section \ref{sec:sr_summary} provides context for the statistical method with a review of engineering models that characterize strike rates to tall structures.  Section \ref{sec:stat_analysis} follows with a mathematical derivation of the method and an example. Section \ref{sec:conus_analysis} applies this technique to a seven-year collection of LLS-detected return strokes near over 67,000 WT locations over the contiguous U.S. (CONUS). The analysis characterizes CG and TUL strike rates and their dependence on total tip height and strike point density. Section \ref{sec:direct_obs} characterizes the variability in strike rates by examining LLS-detected strokes near individual WTs. This analysis serves as both a validation for the statistical method and a pathway to characterize SIUL in cold-season months. Finally, section \ref{sec:summary} offers a summary and concluding remarks. 

\section{Factors affecting lightning strike rates to tall structures}\label{sec:sr_summary}

The total strike count to a WT represents a sum of counts from each of the three aforementioned categories (Note: this paper will use ``UL'' to represent a result that includes both TUL and SIUL, see Section \ref{sec:conus_analysis})

\begin{equation}
N_\mathrm{T}=N_\mathrm{CG} + N_\mathrm{TUL} + N_\mathrm{SIUL}
\label{eq:tot_N}
\end{equation}

\noindent Each count in any given year is highly variable. However, with a sufficiently large sample, we can relate the statistical averages of the number of strikes of each type to features such as the local strike density and total tip height. This section summarizes observations and fitted engineering models that aim to characterize the expected number of strikes from each lightning category. These formulations form the basis for the statistical models introduced in Section \ref{sec:stat_analysis}.

In the absence of terrain variation or structures, the ground strike density is uniform over a sufficiently long time. We refer to this average background density $\rho$ as the local ground strike density. Tall structures, including WTs, promote the development of upward connecting leaders and thus attract nearby stepped leaders that precede CG return strokes. The dynamics of this process can be captured by the concept of a "striking distance" $r_\textrm{s}$, defined as the distance between a downward leader and a structure at which the leader makes contact with the structure. Several theoretical and observational studies have sought to characterize the striking distance, and these findings have been crucial to developing lightning protection standards \citep{golde1945frequency,armstrong1968field,ait2009modified,iec62305}. Modern research on factors controlling lightning attachment to tall structures, such as the dependence on the line charge of the stepped leader, leverages detailed numerical simulations of the various processes involved the attachment process \citep{dellera1990lightning,dellera1990lightningb,li2012physical,becerra2006simplified,becerra2006self,yang2017analysis}, including accounting for the effects of blade rotation \citep{fan2024investigation}. 

The net effect of the striking distance on the CG strike rate is captured as an engineering parameter called the attraction radius $R$ \citep{yang2017analysis}. The area defined by the attraction radius, referred to as the collection area $A$, yields the total number of strikes to the turbine when multiplied by the local ground strike density \citep{eriksson1987incidence,rachidi2008review,international2019wind}:

\begin{equation}
    N_\mathrm{CG} = \underbrace{\left(\pi R^2\right)}_{A} \rho
\label{eq:attraction_radius}
\end{equation}

\noindent Some publications reference the collection area to the ground flash density, and in this work we begin with a stroke density. To avoid confusion we attach a superscript label to attraction radii and density to distinguish between a reference to local ground flash (GF), strike point (SP), and stroke (ST) density. 

As with downward CG strikes, the number of triggered UL strikes $N_\mathrm{TUL}$ scales with the local flash density $\rho$. However, the concept of a collection area does not apply, as TUL is initiated not by an approaching stepped leader but rather by leader development induced by a nearby flash \citep{montana2016lightning}. Since both $N_\textrm{CG}$ and $N_\textrm{TUL}$ are proportional to $\rho$, one can empirically determine the multiplication factor by analyzing the ratio between upward and downward discharges. Using tall tower observations and climatologically based lightning flash incidence rates, \citet{eriksson1987incidence} estimates the probability of upward lightning $P_\textrm{UL}$ as a function of structure height $H$ (m):

\begin{equation}
    P_\textrm{UL} = 52.8\ln(H) - 230
    \label{eq:frac_up_erikson}
\end{equation}
    
\noindent where $H>78$ m.  Equivalently, we can consider an ``upward'' attraction radius $R_\textrm{UL}$ or a total effective attraction radius that includes upward and downward discharges. 

Using an observed correlation between ground flash density and thunderstorm days, along with the observed incidence of strikes to structures of various heights, \citet{eriksson1987incidence} derived the following relationship for the total flash collection area (in m$^2$) as a function of tip height $H$ (m)

\begin{equation}
    A^\mathrm{(GF)}=24H^{2.05}
\label{eq:erikson_afl}
\end{equation}

\noindent The WT lightning protection standard IEC 61400-24 adopts a similar approach but includes a separate location scaling factor $C_\mathrm{D}$ to account for the increased effective attraction radii from local terrain effects and height above sea level \citep{international2019wind}. The IEC 61400-24 standard assumes a strike point-normalized collection area with radius $3H$, giving a total WT strike rate of

\begin{equation}
N = \rho^\mathrm{(SP)} \underbrace{A^\mathrm{(SP)}}_{=\pi (3H)^2} C_\mathrm{D}
\label{eq:iec61400_nd}
\end{equation}

The location factor $C_\mathrm{D}$ also accounts for winter lightning activity, which is dominated by SIUL (see Annex B of \citet{international2019wind}). Since SIUL occurs primarily in cold months and is not associated with normal warm-season thunderstorms \citep{warner2014synoptic,pineda2018thunderstorm}, an effective collection area concept doesn't apply to this category of lightning.  Instead of scaling with an annualized local flash density, the total number of SIUL events $N_\mathrm{SIUL}$ depends on both the total time when conditions are favorable for upward lightning and the likelihood of an SIUL discharge, which is influenced by the height and exposure of the structure \citep{rizk1994modeling,becerra2018estimation}. \citet{montanya2016global} illustrates where these favorable conditions exist based on observed winter lightning trends. However, since tall structures themselves are a source of SIUL, an analysis of historical winter lightning climatology patterns may be insufficient to determine where favorable conditions exist given the presence of a tall structure.

The statistical analysis approach detailed in Section \ref{sec:stat_analysis} assumes the strike rate for both the upward and downward discharges is proportional to the local ground strike density $\rho^\textrm{(SP)}$, and evaluates an effective attraction radius for each type. That is, the method targets statistics for CG and TUL strike rates. The occurrence of SIUL is addressed in Section \ref{sec:direct_obs}.

\section{Statistical Analysis Method}\label{sec:stat_analysis}

The analysis begins with constructing a histogram of LLS-detected stroke counts binned by distance from each turbine.  The total observed counts represent a superposition of downward CG strokes and upward flashes. The CG strokes are seen to preferentially collect on the turbine location due to the striking radius effect, creating a well of lower density immediately surrounding the turbine. In the absence of upward flashes, the observed collection of strokes hitting the turbine should be compensated by this deficit. However, upward flashes represent an \emph{additional} population of discharges that would not have occurred without the turbine's presence. Therefore, in cases where measurable upward discharge activity exists, we expect to see a population of strikes to the turbine that exceeds what we would predict based on the deficit in the well.

We start by assuming the annual local CG stroke density is approximately uniform over a given area and is equal to $\rho_0$ strokes-km$^{-2}$. Over reasonably flat terrain, on average the number of observed CG strokes per year in a given area $A$ would be $\rho_0 A$. The presence of a WT disrupts this uniform distribution. Consider a CG stroke that would have hit the ground at a distance $d\!=\!r$ from the WT if it were not installed. After installation, we can hypothesize that the probability of the same stroke hitting the WT instead of the ground is

\begin{equation}
p(\mathrm{strike}|d=r) = \mathrm{e}^{-r/\lambda}
\label{eq:prob_strike}
\end{equation}

\noindent  We hypothesize an exponential decay function because it provides a convenient single-parameter function that captures influence over a given distance. This assumption models the observations quite well as shown below. Future iterations of this method may consider a refined probability function. 

With this assumption, at distances $r\!>\!0$ the lightning stroke density $\rho(r)$ is a well centered at the origin with density

\begin{equation}
\rho(r) = \rho_0(1 - \mathrm{e}^{-r/\lambda}), \,\,{r>0}
\label{eq:well_distn}
\end{equation}

\noindent such that $\rho(r)$ approaches the local background density $\rho_0$ at $r\gg \lambda$. The total number of downward CG strokes that strike the turbine, that is, the strokes attracted to the turbine with probability dictated by \eqref{eq:prob_strike}, is 

\begin{equation}
N_\mathrm{CG} = \rho_0\int_{\theta = 0}^{2\pi}\int_{r = 0}^\infty \mathrm{e}^{-r/\lambda}rdrd\theta = \rho_0 2\pi \lambda^2
\label{eq:tot_strokes}
\end{equation}

\noindent Dividing $N_\mathrm{CG}$ by $\rho_0$ gives the CG collection area in terms of the exponential parameter $\lambda$. Hence, the CG attraction radius referenced to the local stroke density is

\begin{equation}
R_\mathrm{CG}^\textrm{(ST)} = \sqrt{\frac{2\pi\lambda^2}{\pi}} = \sqrt{2}\lambda 
\end{equation}

Mathematically, we first assume that all these downward strokes hit each turbine at a single point, defined here as the origin. The resulting density function contains a Dirac delta function in the radial dimension divided by $2\pi r$ (so that the integral over the 2D plane of $\delta(r)/(2\pi r)$ equals 1) weighted by the total number of strokes given in \eqref{eq:tot_strokes}. Hence  

\begin{equation}
\rho(r) = \rho_0\left(2\pi\lambda^2\frac{\delta(r)}{2\pi r}  + 1 - \mathrm{e}^{-r/\lambda}\right),\,\,r\ge0
\end{equation}

\noindent We now suppose we also have a population of (triggered) upward discharges, and the ratio of upward to downward discharges is $\beta$. Again assuming all strikes occur at the origin, we have

\begin{equation}
\rho(r) = \rho_0\left((1+\beta)\lambda^2\frac{\delta(r)}{r} + 1 - \mathrm{e}^{-r/\lambda}\right),\,\,r\ge0
\label{eqn:density_no_sigma}
\end{equation}

Lightning solutions from an LLS have location uncertainty defined by an error ellipse \citep{cummins1998combined}. For a mathematically tractable solution, we assume that the LLS produces a circularly symmetric location uncertainty that follows a 2D Gaussian distribution with a root-mean-square error $\sigma$ in any direction. With this assumption, a histogram of LLS-located solutions of events that are all centered at the origin follows a probability density function along the radial axis $r=|\mathbf{r}-\mathbf{r}_0|$ that is given by the Rayleigh distribution 

\begin{equation}
f_\mathrm{R}(r) = \frac{r}{\sigma^2}\mathrm{e}^{\frac{-r^2}{2\sigma^2}}
\end{equation}

\begin{figure}
    \centering
    \includegraphics{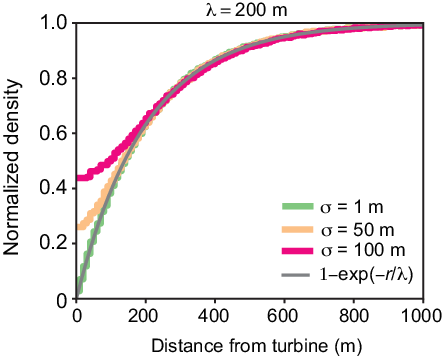}
    \caption{Monte Carlo analysis of the effect of location uncertainty on the stroke density \eqref{eq:well_distn}. The location uncertainty $\sigma$ is shown in the legend.  }
    \label{fig:monte_carlo_well}
\end{figure}

The convolution of $f_\mathrm{R}(r)$ with the Dirac delta term in \eqref{eqn:density_no_sigma} yields the expected distance-dependent density due to strikes at the WT (i.e., the origin):

\begin{equation}
\rho_\mathrm{strike}(r) = \rho_0(1+\beta)\frac{\lambda^2}{\sigma^2}\mathrm{exp}\left(\frac{-r^2}{2\sigma^2}\right)
\label{eq:density_origin}
\end{equation}

\noindent A functional form for the convolution between the Gaussian location uncertainty associated with the LLS and the density well \eqref{eq:well_distn} is less straightforward. Fig. \ref{fig:monte_carlo_well} shows a Monte Carlo simulation of the distribution of events that do not hit the turbine when subjected to a location uncertainty characterized by $\sigma$. The distribution obeys \eqref{eq:well_distn} beyond $r\simeq 2\sigma$, and it flattens out to a minimum value approximately given by $1-\mathrm{exp}(\sigma/\lambda)$. Since the density peak resulting from lightning strikes to the turbine is significantly higher than 1, we assume that the smearing effect from the strokes that do not hit the turbine on the overall distribution is minimal. Thus, the overall density distribution is approximately

\begin{equation}
    \rho(r) \simeq \rho_0\left[(1+\beta)\frac{\lambda^2}{\sigma^2}\mathrm{e}^{\frac{-r^2}{2\sigma^2}} + 1 - \mathrm{e}^{-r/\lambda} \right],\,\,r \ge 0
    \label{eq:tot_density}
\end{equation}

The density can be converted to the number of expected lightning counts versus distance. Using bin size $\Delta r$, the total number of events in a ring centered at $r$ with width $\Delta r$ is

\begin{equation}
    N(r) \simeq 2\pi \rho(r) r \Delta r
    \label{eqn:ltg_counts_vs_dist}
\end{equation}

\noindent The analysis below fits the functional form of lightning counts \eqref{eqn:ltg_counts_vs_dist} to observed distributions of lightning near turbines using the Levenberg-Marquardt algorithm, implemented in the Python package SciPy \citep{Kaltenbach2022Leven-59650}. The parameters of the fit\textemdash $\beta$, $\sigma$, and $\lambda$\textemdash yield the ratio $\beta$ of upward to downward discharges, an approximation $\sigma$ of the LLS location uncertainty, and the total collection area of upward and downward discharges

\begin{equation}
    A_\textrm{T} = 2\pi\lambda^2 (1 + \beta)
    \label{eq:tot_coll_area}
\end{equation}

The total distribution $N(r)$ is accumulated over a total of $N_\textrm{t}$ turbines distributed over the US, using a one-to-many association between LLS-observed strokes and WT locations. That is, $N_i = \sum_j^{N_\mathrm{t}}N_{ij}$. Because $N_\textrm{t}$ is large and the local stroke densities $\rho_0$ vary widely, we normalize to model the shape of the overall distribution, not its absolute value.

\subsection{Density redistribution}\label{sec:density_redistirbution}

Nearby WTs can significantly influence the distribution of lightning counts ${N_i}$. The collection area and upward discharges from neighboring WTs result in an apparent density enhancement at any given turbine, which is most prominent around the typical turbine separation distance. This enhancement introduces perturbations in the histogram, caused by the one-to-many association used to build the overall distribtuion, and it leads to inaccuracies in the fit parameters $(\beta,\sigma,\lambda)$. We propose a density redistribution that partially mitigates the perturbations, as follows: For any stroke that contributes to the count of a WT indexed by $k$ but is closer to another WT indexed by $l\ne k$, we adjust the count at WT $k$ by applying a weight based on the inverse of the normalized density $\tilde{\rho}(d) \equiv \rho(d) / \rho_0$, evaluated at the distance $d_l$ to the nearest WT. That is,

\begin{equation}
    w = \frac{1}{\tilde\rho(r=d_l)}
    \label{eq:weight}
\end{equation}

\noindent This weighting “smears out” the effect of WT $l$ (both captured downward strokes and upward discharges) that perturbs the count associated with WT $k$. A weight of 1 is applied if WT $k$ is the closest to the stroke.

To understand how this weighting method helps to partially negate the perturbation caused by nearby turbines, consider a small area patch $\Delta A$ near turbine $l$. Without turbine $l$, the accumulated counts at turbine $k$ from strokes within this patch would be $\Delta N = \rho_0 \tilde{\rho}(r=d_k) \Delta A$. However, when turbine $l$ is present, the counts may decrease due to the ``well'' effect near turbine $l$, or they may increase if $d_l$ is close enough to turbine $l$ to fall within the location uncertainty of the LLS. In either case, the presence of turbine $l$ modulates the counts at turbine $k$, and this interaction introduces a multiplicative effect: With the nearby turbine $l$, the observed counts at turbine $k$ within this patch become $\Delta N = \rho_0 \tilde{\rho}(r=d_k) \tilde{\rho}(r=d_l) \Delta A$. Weighting each count by the factor described in \eqref{eq:weight} helps to mitigate this modulating effect.

\subsection{An example histogram}

\begin{figure}
    \centering
    \includegraphics{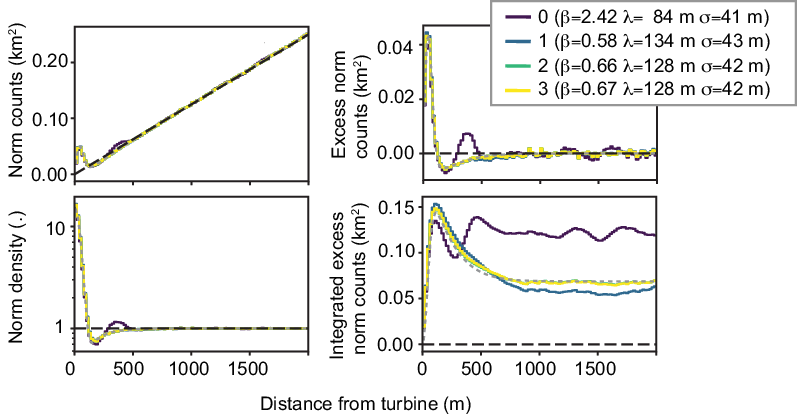}
    \caption{Four representations of a distance-dependent CG stroke count distribution near turbines, where the turbine is at the origin. Color indicates iteration number. The legend at top shows the iteration index and fitted parameter $(\beta,\sigma,\lambda)$ values on each iteration. Dashed gray lines show the fitted curves at the final iteration. Dashed black lines show the expected curve values in the absence of a turbine. See text for a detailed description of each plot.}
    \label{fig:sample_iteration}
\end{figure}

Fig. \ref{fig:sample_iteration} illustrates the normalization process, which is applied iteratively with all weights initialized to 1. After a parameterized curve is fitted to $\{N_i\}$ to determine the parameters $(\beta,\sigma,\lambda)$, the weights for stroke closer to another turbine are determined using \eqref{eq:weight} with $\tilde \rho(r)$ derived from \eqref{eq:tot_density} and \eqref{eqn:ltg_counts_vs_dist}. The iteration process repeats until the fitted count curve converges. The four panels in the figure show different formulations of the same distribution with four iterations of the normalization. The top-left panel displays the total histogram counts $\{N_i\}$ divided by $\rho_0$, where the background density is found by fitting the curve defined by \eqref{eqn:ltg_counts_vs_dist} and \eqref{eq:tot_density} to the observed histogram. The resulting normalized curve has units of area and converges at $d\gg \lambda$ to the area in each distance bin of width $\Delta d = 20$ m. Consequently, the asymptote of the curve reaches $A=2\pi d \Delta d = 2\pi (2) (0.02) \simeq 0.25$ km$^2$ at the largest radius $d=2$ km. 

The other three panels are derived from the normalized count histogram. The bottom-left panel shows the density, where each value is normalized by the area in that bin, i.e., $\tilde \rho_i = N_i / (A_i \rho_0)$, where $A_i \simeq 2 \pi d_i \Delta d$. Normalizing by $\rho_0$ ensures that the density approaches 1 for $d\gg \lambda$. The top-right panel shows the normalized counts with the area subtracted. This plot shows the normalized counts in excess or deficit compared to the counts we would expect if the density were uniform with respect to distance from each turbine, i.e., $A^\mathrm{surplus}_i = N_i/\rho_0 - A_i$.  Finally, the bottom-right panel shows the cumulative summation of $A_\mathrm{surplus}$.

The dashed gray lines in each plot show the best-fit curves obtained from the final iteration. Each equation is straightforwardly derived from \eqref{eqn:ltg_counts_vs_dist} and \eqref{eq:tot_density}.  The lower right panel is derived by subtracting the area $2\pi r dr$ from \eqref{eqn:ltg_counts_vs_dist} and integrating to find

\begin{equation}
        A_\mathrm{cumulative\,\,surplus}(r) = 2\pi\int_0^{r} (\rho(r^\prime)-1) r^\prime dr^\prime
    \simeq 2\pi\lambda^2 \left[(1+\beta)(1-\mathrm{e}^{-r^2/2\sigma^2}) +\mathrm{e}^{-r/\lambda}(\frac{r}{\lambda} + 1)-1  \right]
        \label{eq:integrated_area}
\end{equation}
  
Each iteration involves fitting the parameters $(\beta,\sigma,\lambda)$. These parameters define $\tilde \rho(r)$ for the weights in the next iteration, as described in Section \ref{sec:density_redistirbution}. The legend shows the progression of the three parameters on each iteration and the iteration number. Hence, in all four panels the line marked with index ``0'' reflects the raw normalized counts. Through trial and error, we found that three iterations are typically sufficient for convergence.

The bottom-left normalized density plot offers an intuitive understanding of the distribution. The density increases sharply near $r=0$ from the capture of downward CG strokes (the collection area effect) and upward discharges. At $r\gg \lambda$ the normalized density settles to $\sim$1, the normalized average stroke rate. At a distance slightly larger than  $\sigma$, the density dips below 1. This region reflects a depression in the stroke density compared to the background rate, as some strokes that would have contacted the ground nearby instead strike the turbine. That effect decreases with distance with an e-folding scale $\lambda$ via \eqref{eq:prob_strike}. At $\sim$400 m, the density rises in response to contributions from neighboring turbines. This secondary peak in stroke density affects the parameterized fit. However, as iteration number increases, the secondary peak diminishes, and the three fit parameters $(\beta,\sigma,\lambda)$ stabilize after the second iteration. 

The top-right plot showing the normalized surplus count reveals a similar pattern, but amplifies features at larger distances as the integration area scales with the bin radius. This distribution exhibits noticeable secondary oscillations around 1200 and 1800 m, again caused by neighboring turbines. This secondary oscillation is largely removed by normalizing the counts by the fitted distribution in each iteration.

The bottom-right plot showing the cumulative surplus counts offers a clear interpretation of the collection area and the ratio between upward and downward strikes. Since the fit normalizes the surplus counts to $\rho_0$, the y-axis is expressed in units of area. If $\sigma=0$, then this plot would integrate a Dirac delta function and immediately rise to a peak representing the total collection area.  However, with $\sigma>0$, the integrated peak is smoothed out and therefore does not rise to the total collection area value. The integration of the smeared-out peak also interacts with the well region where there is a deficit of downward strokes due to the capture by the turbine. If the turbine generated no upward lightning, then the cumulative curve would approach zero where $r\gg \lambda$, but as shown, the actual curve settles at a non-zero value.  With reference to \eqref{eq:integrated_area}, this value simply equals the effective upward discharge collection area, i.e., $A_\mathrm{cumulative\,\,surplus}(r)\rightarrow 2\pi\lambda^2 \beta$. 

This statistical estimate of upward discharge rates suffers from an obvious limitation that the LLS misses a significant fraction of actual upward discharges. However, since this technique separates the population of upward and downward discharges, a scaling factor $\xi$ can be applied to the upward discharge count based on an assumed upward discharge detection efficiency, thereby statistically correcting for the UL under-counting.

\section{CONUS-wide statistical analysis of strokes near wind turbines}\label{sec:conus_analysis}

\begin{figure}
    \centering
    \includegraphics{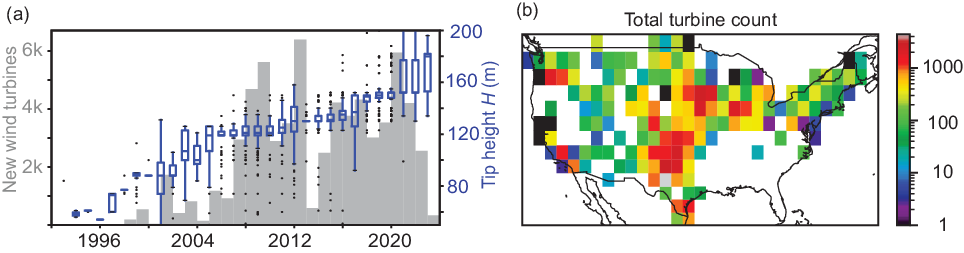}
    \caption{(a) New wind turbine operational count (gray histogram) and total tip height distribution (box and whiskers plot). Box plot whiskers set at 1.5 times the interquartile range beyond the 25$^\textrm{th}$ and 75$^\textrm{th}$ quartiles. (b) Total number of WTs, accumulated over a 2$\times$2 degree grid.} 
    \label{fig:wt_height}
\end{figure}

We present an application of the statistical fitting method detailed above using seven years (2017-2023 inclusive) of CG stroke data from the NLDN together with WT data reported in the US Wind Turbine Database (USWTDB v6.1, as released on 2023-11-28; \citep{wswtd}). The NLDN data have been reprocessed using the latest (2021 release) version of the classification algorithm \citep{murphy2021recent,murphy2021may}. The USWTDB contains 27 fields, of which we use five: latitude, longitude, the year the project became operational, turbine total height (maximum tip height $H$), and location confidence (reported from 0-3). Only WT entries with the highest location confidence are included, 67,565 in total. Fig. \ref{fig:wt_height}a shows the number of WTs that became operational since 1990, together with a box and whisker plot characterizing the distribution of total tip heights by year. Fig. \ref{fig:wt_height}b shows how these turbines are distributed across CONUS. 

The analysis begins by matching all CG strokes reported by the NLDN over the seven-year period to the locations of the 67,565 WTs with high location confidence. Each stroke is matched to all WTs within 2 km that were operational at the time of the stroke, ensuring the turbine was present. This process results in 24,512,736 WT-stroke matches. For each stroke matched to WT $k$, the distance to WT $d_k$ is recorded, as well as the distance to the closest WT, if different. The closest distance, when shorter than distance $d_k$, controls the density redistribution weighting using \eqref{eq:weight}.

Distance-dependent histograms can be constructed using subsets of the matched data to explore the dependence of strike rates on either WT features (e.g., height) or stroke characteristics (e.g. season). Each distribution uses a distance bin size of 20 m, ranging from 0 to 2 km. Iterative functional fits applied to each distribution yield the parameters $(\beta,\lambda,\sigma)$, which, in turn, determine the downward and upward attraction radii and the LLS uncertainty. Each fit undergoes three iterations.

\begin{figure}
    \centering
    \includegraphics{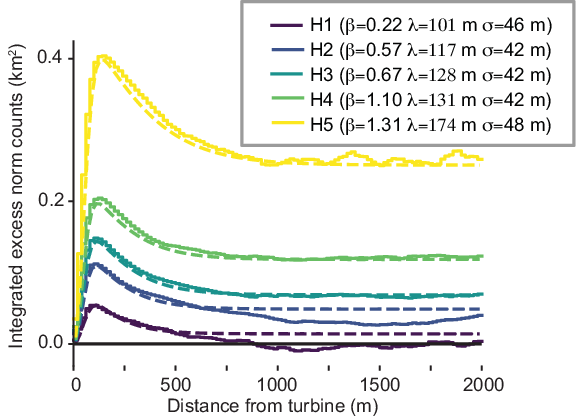}
    \caption{Integrated excess normalized counts for five categories of total tip heights. }
    \label{fig:integrated_by_height}
\end{figure}

The results presented here consider five categories of tip heights, labeled as follows: H1 (85-115 m), H2 (115-130 m), H3 (130-140 m), H4 (145-160 m), and H5 (160-200 m). These height ranges are selected to ensure sufficient statistical representation in each category. Fig. \ref{fig:integrated_by_height} displays the final iteration of the integrated excess normalized counts (similar to the bottom-right panel in Fig. \ref{fig:sample_iteration}) for each of these tip height categories, along with the three statistical parameter fits of $(\beta,\lambda,\sigma)$ from the final iteration. The dashed lines show the fitted results applied to \eqref{eq:integrated_area}. The close correspondence between the observed cumulative distribution and the fitted parameters reinforces the reliability of the analysis technique and the derived parameters.

The progression of the three fit parameters with height reveals several key insights into the interaction between lightning and turbines as a function of tip height.  First, the fitted $\sigma$ parameter remains relatively constant at $\sigma\simeq 45$ m. This consistency indicates that the LLS's precision in locating lightning attachments to turbines is largely invariant with tip height. In contrast, the $\lambda$ parameter, related to the downward stroke collection area through $A_\mathrm{CG} = 2\pi\lambda^2$, shows a strong dependence on tip height. The taller tip heights have a larger striking radius, attracting downward CG strokes over a larger effective collection area. 

The upward to downward discharges ratio also increases with height, from a relatively low ratio of $\beta=0.22$ in the lowest tip height category to $\beta=1.31$ for tip heights above 160 m. Since the upward collection area is $A_\mathrm{UL} = 2\pi\lambda^2 \beta$, the rate of upward discharges also increases with height and, when normalized by the local CG stroke density, increases at a faster rate compared to WT strikes by downward CG strokes. 

Fig. \ref{fig:integrated_by_height} derives the fit parameters by normalizing to the local \emph{stroke} density $\rho_0$. The $\lambda$ and $\beta$ parameters yield the downward and upward stroke attraction radii via, respectively

\begin{equation}
    \begin{array}{ll}
        R_\textrm{CG}^\textrm{(ST)} &= \sqrt{2}\lambda \\
        R_\textrm{UL}^\textrm{(ST)} &= \sqrt{2\beta}\lambda
    \end{array}
\label{eq:attraction_radii_lambda}
\end{equation}

\noindent For a direct comparison against the IEC 61400-24 standard, we aim to normalize the attraction radii to a local strike point density. Assuming a conversion between the ground stroke and strike point densities $\rho^\textrm{(ST)} = \alpha^\textrm{(SPST)}\rho^\textrm{(SP)}$, and referring to the number of WT strikes, the radii are related as follows:

\begin{equation}
    N = \pi (R^{\textrm{(ST)}})^2\rho^\textrm{(ST)} = \pi (R^{\textrm{(SP)}})^2\rho^\textrm{(SP)} 
    \rightarrow \, R^{\textrm{(SP)}} = \sqrt{\alpha^\textrm{(SPST)}}R^{\textrm{(ST)}}
\label{eq:st_to_sp_conversion}
\end{equation}

\noindent Similarly, to convert a flash-referenced attraction radius $R^\textrm{(GF)}$ to a strike-point-referenced radius $R^\textrm{(SP)}$, assuming a density conversion $\rho^\textrm{(SP)} = \alpha^\textrm{(GFSP)}\rho^\textrm{(GF)}$, we have

\begin{equation}
    R^{\textrm{(SP)}} = \frac{1}{\sqrt{\alpha^\textrm{(GFSP)}}}R^{\textrm{(GF)}}
    \label{eq:fl_to_sp_conversion}
\end{equation}

\noindent These conversion factors can be applied to \eqref{eq:attraction_radii_lambda} to convert between stroke, strike point, and flash attraction radii. \citet{vagasky2024} present a CONUS-wide analysis of CG strokes, flashes, and strike points from the NLDN, showing an average of 2.7 strokes per flash, and $\alpha^\textrm{(GFSP)}=1.69$ ground strike points per flash. Therefore, we estimate $\alpha^\textrm{(SPST)} \simeq 2.7/1.69=1.6$ strokes per ground strike point. 

\begin{figure}
    \centering
    \includegraphics{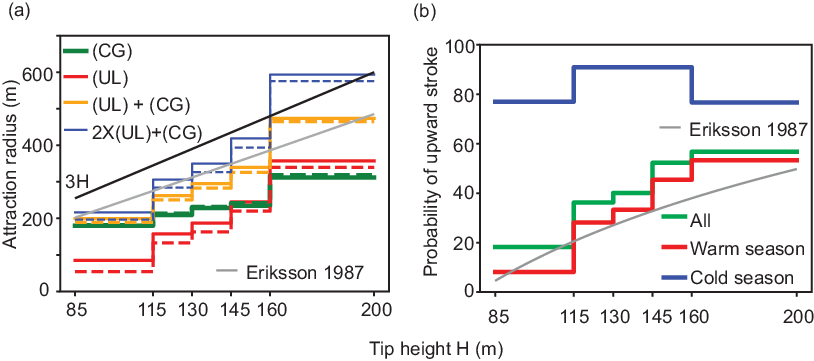}
    \caption{Statistical WT strike analysis segmented by total tip height. (a) Measured strike-point referenced attraction radii for downward (green), upward (red), total (orange), and a scaled total assuming a 50\% upward discharge detection efficiency (blue). Solid color lines show attractive radii using data from all months. Dashed lines show the corresponding radii using only data from the warm thunderstorm season (May through August). The IEC 61400-24 collection radius of $3H$ line and the radius derived from \eqref{eq:erikson_afl} from \citet{eriksson1987incidence} shown as black and gray lines, respectively. (b) Fraction of upward strokes for all seasons (green), warm season (red), and cold season (blue), compared with the estimate from \eqref{eq:frac_up_erikson} in \citet{eriksson1987incidence} (gray).} 
    \label{fig:collection_radii_and_prob_upward}
\end{figure}

Fig. \ref{fig:collection_radii_and_prob_upward}a plots the strike point-referenced upward (solid red) and downward (solid green) attraction radii, calculated by applying \eqref{eq:attraction_radii_lambda} to the $\lambda$ and $\beta$ parameters derived in Fig. \ref{fig:integrated_by_height}. The total effective strike point attraction radius, which includes both upward and downward discharges and is given by  $\sqrt{2\alpha^\textrm{(SPST)}(1+\beta)}\lambda$, is shown in orange. The blue curve shows the collection radius obtained by scaling up the upward discharge collection area by a factor of 2. This scaling reflects the total effective attraction radius under the assumption of 100\% CG stroke detection and 50\% UL detection, which is roughly consistent with the $\sim$43\% upward discharge detection efficiency reported in \citet{diendorfer2015lls}. 

These radii are compared against the ``$3H$'' attraction radius used in the IEC 61400-24 standard (black line) and the attraction radius derived from tower strike observations \eqref{eq:erikson_afl} proposed by \citet{eriksson1987incidence} (gray line). The latter attraction radius has been converted from flash-referenced to strike point-referenced using $\alpha^\textrm{(SPST)}$ as described above. The orange curve, representing the measured total attraction radius, shows remarkable consistency with the height dependence proposed over three decades ago by \citet{eriksson1987incidence}. However, due to the lower DE for UL, this curve underrepresents the total strike counts for WTs. The DE-corrected estimate shown by the blue curve aligns more closely with the ``3H'' line from the IEC61400-24 standard, although the differences could have practical significance. At lower tip heights, the IEC standard may overestimate the average number of strikes, while at the upper extreme, the IEC 61400-24 standard appears to underestimate the total. Given the steeper slope of the UL trend line compared to the CG trend line, this discrepancy is likely to worsen as tip heights increase, largely from the disproportionately high number of UL discharges.

The solid curves in Fig. \ref{fig:collection_radii_and_prob_upward}a and the preceding discussion grouped all UL discharges into a single category, meaning that the upward attraction radii encompass both TUL and SIUL. However, since only TUL is expected to depend on the local ground strike density during the primary thunderstorm season, it is more appropriate to filter the data to focus on the warm convective season months. We crudely define the primary thunderstorm season as spanning four months: May through August, inclusive, and refer to this period as the ``warm season''. The dashed lines in Fig. \ref{fig:collection_radii_and_prob_upward}a represent the resulting attraction radii using LLS data filtered for the warm season months. While the downward (CG) attraction radius (green) shows little change, the attraction radius for upward lightning (orange, blue, red) decreases across all tip height ranges. 

Assuming that all UL during the warm season is attributable to TUL, we can estimate a simple regression fit from the dashed curves as follows:

\begin{subequations}
    \begin{align}
            R^\textrm{(SP)}_\textrm{CG} &= 11 + 1.6H \\
            R^\textrm{(SP)}_\textrm{TUL} &= -300 + 3.5H 
    \end{align}
    \label{eq:tot_strikes_wt_conclusion}
\end{subequations}

\noindent where the input tip heights to the regression fit are taken as the midpoint of each tip height range.  The attraction radius for downward CGs intersects the $H$-axis near the origin and has a modest slope of 1.6. In contrast, the TUL attraction radius crosses the $H$-axis at 85 m, close to the 78 m threshold implied by \eqref{eq:frac_up_erikson}. The slope of the TUL attraction radius is notably higher at 3.5. This steeper slope suggests that as tip heights increase, the ratio of upward to downward discharges will likely continue to grow, consistent with the observations reported by \citet{eriksson1987incidence}.

A direct comparison to \eqref{eq:frac_up_erikson} can be made by examining the attraction areas for upward versus all discharges, i.e., $\beta / (1 + \beta)$. Fig. \ref{fig:collection_radii_and_prob_upward}b shows this ratio for all season (green), warm season only (red), and for a ``cold season'' lightning only, defined here as the four months between November through February, inclusive. The warm season-only result, which more accurately reflects the ratio between CG and TUL during periods of normal thunderstorm activity, shows a reasonable agreement with the tower-derived formula from \citet{eriksson1987incidence}.  In contrast, the cold season fraction upward plot is significantly higher than the warm season line and does not exhibit a clear trend with tip height. This result supports the view that cold season lightning has a much higher fraction of SIUL, and hence the total strike rate is not directly proportional to the nearby downward strike rate. In other words, the concentration of lightning near the WT during the cold season is not balanced by a nearby well of attracted CG strokes.

\section{Validation and discussion using a per-turbine strike analysis}\label{sec:direct_obs}

The attraction radii calculated in Section \ref{sec:conus_analysis} only represent \emph{average} values. The statistical fit method does not provide information on the variation among the turbines, nor does it pinpoint which specific strokes struck a turbine. This section uses a distance proximity threshold to estimate the number of strokes that hit individual WTs. The resulting analysis provides a validation of the statistical method, and enables a characterization of the variability in attraction radii among WTs in a given height range. The per-turbine analysis also helps us characterize strike rates in cold-season months, which are, in many regions, dominated by SIUL. 

To identify strokes likely to have struck the turbine, we aim to define a radius, $d_0$, such that if the distance between a  LLS-detected stroke and the WT is less than $d_0$, the stroke is tagged as a probable turbine attachment. The choice of $d_0$ involves balancing the risk of missing an actual attachment stroke against the risk of including nearby ground strokes that did not attach to the turbine. By assuming a Rayleigh distribution of scatter around each WT and that the attachment point is precisely at the mast location, the observed LLS uncertainty $\sigma$ from the statistical analysis gives the fraction of true WT strikes properly identified for a given $d_0$. Simultaneously, assuming a stroke density near each turbine given by \eqref{eq:well_distn}, the integrated average count of non-attachment strokes that fall within the radius $d_0$ is $\rho \pi[d_0^2 + 2\lambda\textrm{e}^{-d_0/\lambda}(d_0 + \lambda) - 2\lambda^2]$.  This count should be evaluated as a fraction of the total number of WT strikes, derived from \eqref{eq:attraction_radius} and \eqref{eq:tot_coll_area} as $\rho \pi 2\lambda^2(1+\beta)$.

From Fig. \ref{fig:integrated_by_height}, we observe that the LLS uncertainty is $\sim$45 m across all WT tip height ranges. Using $d_0=2\sigma\sim 90$ m gives a capture of $\sim$86\% of all strokes attaching to the WT. Using the $\lambda$ and $\beta$ values from the same figure, this choice of $d_0$ results in an upper bound fraction of strokes counted by chance ranging from 14\% for the lowest tip height range to 2\% for the highest tip height range. Consequently, we expect this choice of threshold to give a good approximation of the number of strikes for the shortest tip height range and underestimate the strike count for the tallest range.

This proximity approach also allows us to estimate the attraction radius for \emph{each turbine} by taking the ratio of the local stroke density to the number of strokes within $d_0$ of the WT. The local stroke density for a given turbine is determined by counting the number of strokes in the annulus defined by the distance range of 1 and 2 km. To minimize the influence of nearby WTs, any strokes within $d_0$ of another WT are excluded from the background stroke count. Denoting the local stroke density determined at each WT as $\rho^\textrm{(ST)}_{1-2\,\textrm{km}}$ and using $d_0=90$ m, the total effective attraction radius (including both upward and downward discharges) is then:

\begin{equation}
    R_\textrm{T}^\textrm{(SP)} = \sqrt{\alpha^\textrm{(SPST)}}\sqrt{\frac{N_{90\,\textrm{m}}}{\pi\rho^\textrm{(ST)}_{1-2\,\textrm{km}}}}
    \label{eq:per_turbine_R}
\end{equation}

\begin{figure}
    \centering
    \includegraphics{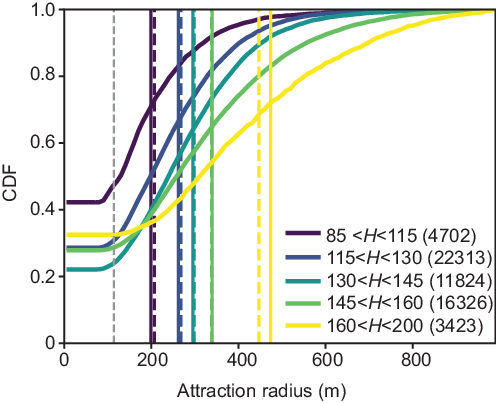}
    \caption{Cumulative distribution function (CDF) of per-WT strike point attraction radius for five categories of WT heights marked in the legend. Counts in parentheses indicate the number turbines in each category. Solid vertical lines indicate the total collection radii from the functional fit from Fig.~\ref{fig:collection_radii_and_prob_upward}a. Dashed vertical lines mark the average attraction radius $R$ calculated from the proximity method. The dashed gray line shows the attraction radius corresponding to random collection in the 90 m inner circle. See text for additional details.}
    \label{fig:cdf}
\end{figure}

Fig. \ref{fig:cdf} plots the cumulative distribution of \eqref{eq:per_turbine_R} across the five categories of total tip height. These distributions provide insight into the significant variability in strike rates to WTs within each height category. While the attraction radii skew to higher values for taller tip heights, $R_\textrm{T}$ ranges from 0 to approximately twice the average value within each category. The concentration of the distribution at $R_\textrm{T}=0$ represents WTs that had no detected strikes during the analysis period; $\sim$20\% to $\sim$37\% of turbines are in that category. The fraction with zero counts decreases with height across the first three height categories but surprisingly reverses for the top two categories. Since the taller height categories correspond to newer WT installations (as shown in Fig.~\ref{fig:wt_height}), a likely explanation for this reversal is the shorter accumulated analysis period for these turbines, which increases the likelihood of observing zero strikes.

The cumulative distribution for each category begins to rise from the zero-strike level at a distance threshold that approximately aligns with the random accumulation of strikes within the $d_0=90$ m strike radius. That is, assuming a uniform local density $\rho$, then 

\begin{equation}
    R_\textrm{random}^\textrm{(SP)} = \sqrt{\alpha^\textrm{(SPST)}}\sqrt{\frac{\pi 90^2 \rho}{\pi\rho}} = 1.26\times 90 = 113\,\textrm{m}
    \label{eq:per_turbine_R_random}
\end{equation}

\noindent This distance is shown with the dashed gray line. Fig. \ref{fig:cdf} also includes vertical solid lines indicating the attraction radius determined from the functional fit for the entire distribution, shown by the orange curve in Fig.~\ref{fig:collection_radii_and_prob_upward}a.  In each category, the attraction radius from the fitted curve intersects the distribution at roughly the 70$^\textrm{th}$ percentile. This implies that roughly one-third of WTs have total attraction radii larger than the averages shown in Fig.~\ref{fig:collection_radii_and_prob_upward}a.

Finally, the dashed lines in the figure show the average total collection radius based on observed strokes within $d_0$ of each turbine. This average is again derived using \eqref{eq:per_turbine_R}, but with $N_{90\,\textrm{m}}$ and $\rho_{1-2\,\textrm{km}}$ replaced by the accumulation of WT-strikes and nearby strokes within 1-2 km, respectively, across all turbines in each height category. As expected from the arguments above, with this choice of $d_0$ the proximity method provides a close estimate of the number of strikes for the lowest tip height range (i.e., the attraction radius determined by counting strokes within $d_0$ of each turbine is close to the fitted value) and underestimates the number for the highest tip height range. 

Several factors contribute to the variability in total effective attraction radii among individual turbines. One factor is statistical variation from low local density and/or limited number of years in the analysis period. Other potentially influential factors, which we leave for future work to fully characterize, include the increased effective height due to terrain and the shielding effects based on the turbine's position within a wind farm. 

\subsection{Cold-season lightning}\label{sec:cold_season}

\begin{figure}
    \centering 
    \includegraphics{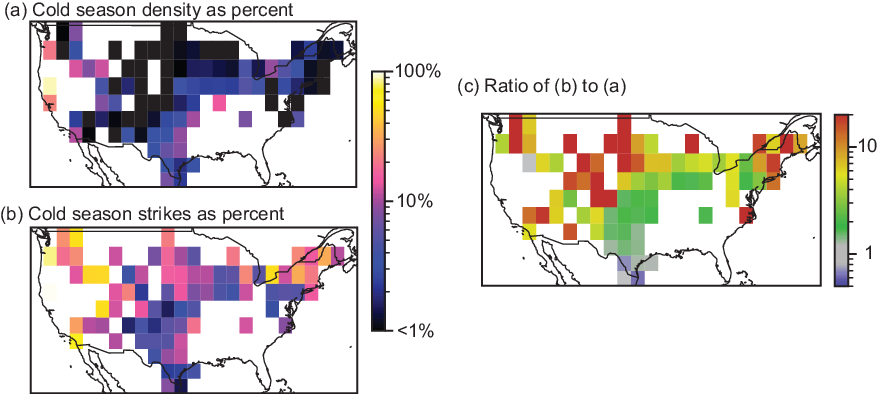}
    \caption{Regional analysis of cold-season per-turbine stroke statistics over a 2.5$\times$2.5 degree grid. (a) Cold-season local density is expressed as a fraction across warm and cold seasons. (b) Fraction of WT strikes observed during the cold season, compared to the total strikes across both seasons. (c) The ratio of the values in panel b to those in panel a, using a color scale ranging from 0.5 to 20. Panels b and c only include pixels with at least 80 accumulated observations years across turbines within that pixel.}
    \label{fig:region_analysis_cold}
\end{figure}

The inclusion of seasons and regions with SIUL also contributes to the variability in per-turbine attraction radii since the effective collection area concept doesn't apply to this category of lightning strikes. We explore this point with Fig.~\ref{fig:region_analysis_cold}, which shows regional patterns of the percent of local CG strike point density and WT-strike counts (determined using the proximity method) during the cold season months (November through February). Fig.~\ref{fig:region_analysis_cold}a shows the cold season local strike point density as a percent of the total local density integrated across both the warm and cold seasons (8 months total). The cold-season local density fraction is less than 1\% over much of the U.S., and rises to over $\sim$10\% near North Texas and in parts of the Mountain West. Fig.~\ref{fig:region_analysis_cold}b shows the percent of cold-season WT strikes relative to the total strike count across both the warm and cold seasons for all tip heights. 

Fig.~\ref{fig:region_analysis_cold}c plots the ratio between panels b and a. The spatial patterns in this ratio demonstrate the likely regionally-variable impact of SIUL on overall strike rates. If cold-season strike rates were driven by the same lightning interaction mechanisms that dominate in the warm season\textemdash specifically, the attraction radius for downward CGs and the effective attraction radius for TUL\textemdash we would expect the ratio of cold-season strike rates to cold-season density to be consistent. In contrast, in regions  dominated by SIUL, which can occur without nearby lightning, we expect the ratio of cold-season WT strike rates to exceed that of the local cold-season CG strike point density.  In southern Teaxs, the ratio is close to or slightly less than 1, indicating that cold-season months experience ``normal'' convective thunderstorm activity, where strike counts remain proportional to the local strike point density. In this regime, the attraction radius concept remains applicable. The ratio steadily increases at more northern latitudes, and exceeds 10X in much of the Mountain West, upper Midwest, and Northeast. This increase indicates a much higher rate of cold season WT strikes than would be predicted by the local ground strike point density. These regions are also more likely to have conditions favorable for SIUL during the cold season, suggesting that much of the observed increase in WT strikes can be attributed to SIUL. 

\section{Summary and conclusions}\label{sec:summary}

This paper introduces a statistical method for analyzing LLS data near wind turbines WTs. This approach effectively captures the attractive effect for downward CG strokes and the presence of triggered upward lightning, as both scale with the amount of lightning in the area. While LLSs cannot reliably differentiate between upward and downward discharges on a per-stroke basis, the functional fit used in the statistical analysis method provides this breakdown on a population level by comparing the integrated number of events against the local lightning density. This separation offers two key benefits in assessing the likely risk to WTs from lightning discharges. First, it enables an ``up-scaling'' of observed strikes from a long-range LLS, which typically misses over half of upward discharges. Second, it allows for a better characterization of the distribution of electrical current parameters related to damage by understanding the ratio of upward to downward discharges.

Referring to \eqref{eq:tot_N}, the total number of lightning strikes can be expressed as the summation of downward CG strokes, triggered upward lightning (TUL), and self-initiated upward lightning (SIUL), with each term given by

\begin{subequations}
  \begin{align}
            N_\textrm{CG} &= A^\textrm{(SP)}_\textrm{CG}\rho^\textrm{(SP)} = \pi(11 + 1.6H)^2\rho^\textrm{(SP)} \\
            N_\textrm{TUL} &= A^\textrm{(SP)}_\textrm{TUL}\rho^\textrm{(SP)} = \xi \pi(-300 + 3.5H)^2\rho^\textrm{(SP)} \\
            N_\textrm{SIUL} &= f(H,\textrm{thunderstorm conditions})
    \end{align}
    \label{eq:tot_strikes_wt_conclusion}
\end{subequations}

\noindent  The number of downward CG strikes and triggered UL strikes is proportional to the local strike point density $\rho^\textrm{(SP)}$ and scales with a collection area defined by a height-dependent attraction radius, derived from a regression fit to the empirical values shown in Fig.~\ref{fig:collection_radii_and_prob_upward}a. The scaling factor ``$\xi$'' in the triggered UL expression accounts for the imperfect LLS detection efficiency for upward flashes, particularly those that do not produce return strokes. For precision networks like the NLDN, this scaling factor is likely in the range of 2 to 3. An analogous scaling factor should be applied to the CG strike rate if the CG stroke DE is less than 100\%. A regional analysis of likely warm and cold season lightning strike counts to WTs shows that in northern and coastal regions cold-season strike rates are not proportional to the local strike point density. In these regions, cold-season strikes are likely dominated by SIUL and thus depend on favorable atmospheric conditions that promote upward discharges. 

The statistical method does not measure the variation in strike rates to individual turbines. A per-turbine strike climatology revealed the total attraction radius varies by over a factor of 2 around the average values. Several controlling factors likely contribute to this variation, including terrain effects, shielding within large wind farms, variable contributions from cold season lightning, and insufficient statistics for individual WTs. Further characterization of these features is left for future work. Furthermore, this study did not differentiate between positive and negative lightning, explore the dependence of attraction radii on peak current, or determine the effects of misclassification between cloud and CG strokes in the NLDN. The statistical and analytical methods presented here can be extended to explore additional features that may improve predictive skill for WT strike rates. As the wind energy industry expands with higher tip heights and a larger geographic footprint, further characterizing these factors will become increasingly important.

\bibliographystyle{unsrtnat}
\bibliography{references}

\end{document}